\documentclass []{article}

\newcommand{\tr}{{\rm tr}}

\newcommand\mathC{{\mkern1mu\raise2.2pt\hbox{$\scriptscriptstyle|$}
        {\mkern-7mu\rm C}}}
\newcommand{\mathR}{{\rm I\! R}}                

\renewcommand\mathR{{\rm I\! R}}

\renewcommand\tr{{\rm tr}\,}

\newcommand\unit{{\rm 1\kern-3.2pt I}}

\begin{document}
\title{Histories quantisation of parameterised systems:
I. Development of a general algorithm }
\author{K. Savvidou \thanks{ntina@ic.ac.uk}\\ {\small Theoretical Physics Group, The Blackett
Laboratory,} \\ {\small Imperial College, SW7 2BZ,
London, UK} \\ {\small and} \\ 
{\small Center for Gravitational Physics and Geometry, Department of Physics,}\\
{ \small Pennsylvania State University, University Park, PA16802, USA} \\ \\
C. Anastopoulos \thanks{charis@physics.umd.edu} \\
{\small Department of Physics, University of
Maryland, College Park, MD20742, USA} } 
\maketitle
\begin{abstract}
We develop a new algorithm for the quantisation of
systems with first-class constraints. Our approach
lies within the (History Projection Operator) continuous-time histories
quantisation programme. In particular, the Hamiltonian treatment (either classical or quantum) of parameterised systems is characterised by the loss of the notion of time in the space
of true degrees of freedom ({\em i.e.\/} the `problem of time'). 
The novel temporal structure of the HPO theory (two laws of time transformation that distinguish between the temporal logical structure and the dynamics) persists after the
imposition of the constraints, hence the problem
of time does not arise. We expound the algorithm for both the classical and quantum cases and apply it to simple models. 
\end{abstract}

\renewcommand {\thesection}{\Roman{section}}
 \renewcommand {\theequation}{\thesection. \arabic{equation}}
\let \ssection = \section
\renewcommand{\section}{\setcounter{equation}{0} \ssection}
\pagebreak

\section{Introduction}
\subsection{Preamble}
In this paper we propose an algorithm for both the
classical and quantum treatments of constrained
systems, based on the notion of consistent
histories. The scheme lies within the general
framework of the consistent histories approach to
quantum theory \cite{Gri84,Omn8894,GeHa9093,Har93a}. More
precisely, we generalise and extend the results of
the HPO (History Projection Operator)
continuous-time histories programme \cite{
I94,IL94,IL95,ILSS98,Sav99,An00} to incorporate the quantisation
of systems with first-class constraints. In
particular, we shall see how these methods bring
substantially new insights for dealing with
parameterised systems.

The Hamiltonian treatment of (first-class)
constrained systems is well understood. In the
classical case the aim is to construct the reduced
phase space, {\em i.e.}, a symplectic manifold on
which the true degrees of freedom are defined, and
then consider the time evolution induced by the
Hamiltonian. The quantum treatment can proceed
with reduced phase space quantisation, where one
employs the standard quantisation algorithms for
the classical theory on this space. Alternatively,
one can use a Dirac type of quantisation where the
Hilbert space of the unconstrained system is
constructed and then the constraint is implemented
quantum mechanically by finding the projector onto
the {\em physical subspace\/ } of the Hilbert
space.

However for a particular class of system, namely
the parameterised systems, the canonical treatment
leads to some deep conceptual problems.
Parameterised systems have a vanishing Hamiltonian
when the constraints are imposed. Classically this
implies that the elements of the reduced phase
space are themselves {\em solutions to the
classical equations of motion\/}, hence they
define a classical history of the system. On the
other hand, a point of phase space corresponds to
a possible configuration of the physical system at
an instant of time. As a result, the notion of
time is unclear, or even ambiguous, in these
systems: in particular, it is not obvious how to
recover the notion of temporal ordering unless we
choose to arbitrarily impose a gauge-fixing
condition. This is also true at the quantum level.
There are no dynamics in the physical Hilbert
space because the Hamiltonian operator vanishes;
the system seems to be in a `frozen' state. This
is one facet of the ``problem of time'' for
parameterised systems, and is particularly acute
in the Hamiltonian (or path integral) treatment of
general relativity (see \cite{I92,Kuc91} for a
 thorough review).

 The HPO continuous-time histories algorithm is particularly suited
to deal with systems that have a non-trivial
temporal structure. This is due to the striking
property that HPO histories admit two distinct
laws of time transformation, each representing a
distinct quality of time \cite{Sav99}. The first
corresponds to time considered purely as a
kinematical parameter of a physical system, with
respect to which a history is defined as a
succession of possible events. It is strongly
connected with the temporal-logical structure of
the theory and is related to the view of time as a
parameter that determines the ordering of events.
The second mode in which time is manifested in
this approach, corresponds to the dynamical
evolution generated by the Hamiltonian.
Classically these two laws are nicely intertwined
through the action principle which provides the
paths that are solutions to the classical
equations of motion. For a detailed presentation of the HPO continuous-time programme see \cite{Sav99b}.

The fact that histories describe objects that have
an intrinsic temporality, together with the two
notions of time transformation, enables us to
treat parameterised systems in such a way that the
problem of time does not arise. As we shall see,
HPO histories keep their intrinsic temporality
after the implementation of the constraint, thus
there is no uncertainty about the
temporal-ordering properties of the physical
system.

Another important result is that the theory admits
time reparameterisation as a symmetry: classically
it arises as an invariance of the equations of
motion, while in quantum theory it is an
invariance of the assignments of probabilities.

This has been a very brief description of the aims
and results of this paper. We shall now give some
more details about the underlying concepts and
ideas of the histories quantisation scheme.

\subsection{Consistent histories}
In classical Newtonian theory, time is introduced
as an external parameter; and in all the existing
approaches to quantum theory, the treatment of
time is inherited from the classical theory. On
the other hand, general relativity treats time as
an internal parameter of the theory: in
particular, it is one of the coordinates of the
spacetime manifold. When we attempt to combine the
two theories in quantum gravity, this essential
difference in the treatment of time appears as a
major problem---one of the aspects of what is
known as the `Problem of Time'. One of the
directions towards a solution of the problem is to
construct theories where time is introduced in a
novel way.

One such formalism is the consistent histories
approach to quantum theory in which time appears
as the label on a time-ordered sequence of
projection operators which represents a `history'
of the system. In the original scheme by Gell-Mann
and Hartle \cite{Gri84,Omn8894,GeHa9093}, the crucial
object is the decoherence functional written as
\begin{equation}
d(\alpha,\beta)= {\rm tr}(\tilde
C_\alpha^\dagger\rho \tilde C_\beta)
\label{decfun1}
\end{equation}
where $\rho$
is the initial density-matrix, and where the {\em
class operator\/} $\tilde C_\alpha$ is defined in
terms of the standard Schr\"odinger-picture
projection operators $\alpha_{t_i}$ as
\begin{equation}
 \tilde C_\alpha:=U(t_0,t_1)\alpha_{t_1} U(t_1,t_2)
    \alpha_{t_2}\ldots U(t_{n-1},t_n)\alpha_{t_n}U(t_n,t_0)
\end{equation}
where $U(t,t')=e^{-i(t-t')H/}$ is the unitary
time-evolution operator from time $t$ to $t'$.
Each projection operator $\alpha_{t_i}$ represents
a proposition about the system at time $t_i$, and
the class operator $\tilde C_\alpha$ represents
the composite history proposition ``$\alpha_{t_1}$
is true at time $t_1$, and then $\alpha_{t_2}$ is
true at time $t_2$, and then \ldots, and then
$\alpha_{t_n}$ is true at time $t_n$''.

The consistent histories approach allows the
description of an approximately classical domain
emerging from the macroscopic behaviour of a
closed physical system, as well as its microscopic
properties, in terms of the conventional
Copenhagen quantum mechanics. This is possible
through the decoherence condition: the requirement
for `decoherence' (negligible interference between
disjoint histories) selects a consistent set of
histories that can be represented on a classical
(Boolean) lattice, thus having a classical logical
structure. Hence in the consistent histories
theory, emphasis is given to the observation that,
although in atomic scales a system is described by
quantum mechanics, it may also be described by
classical mechanics and ordinary logic. Therefore
a more refined logical structure seems to be a
necessary part of any consistent histories
formalism. However, the Gell-Mann and Hartle
approach lacks the logical structure of standard
quantum mechanics in the sense that the
fundamental entity ({\em i.e.}, history) for the
description of the system is not represented by a
projector in the Hilbert space of the standard
theory. This is because, as a product of
(generically, non-commuting) projection operators,
the class-operator $\tilde C_\alpha$, that
represents a history, is not itself a projector.

\subsection{The History Projection Operator Approach}
 The difference between the representation of
propositions in standard quantum mechanics and in
the history theory is resolved in the alternative
approach of the `History Projection Operator'
theory \cite{I94,IL94}, in which the history
proposition ``$\alpha_{t_1}$ is true at time
$t_1$, and then $\alpha_{t_2}$ is true at time
$t_2$, and then \ldots, and then $\alpha_{t_n}$ is
true at time $t_n$'' is represented by the {\em
tensor product\/}
$\alpha_{t_1}\otimes\alpha_{t_2}\otimes
\cdots\otimes\alpha_{t_n}$ which, unlike $\tilde
C_\alpha$, {\em is\/} a genuine projection
operator, albeit one that is defined on the tensor
product of copies of the standard Hilbert space
${\cal V}_n = {\cal H}_{t_1}\otimes{\cal
H}_{t_2}\otimes\cdots\otimes{\cal H}_{t_n}$. Hence
the `History Projection Operator' formalism
extends to multiple times, the quantum logic of
single-time quantum theory.

However, the introduction of the tensor product
${\cal H}_{t_1}\otimes{\cal
H}_{t_2}\otimes\cdots\otimes{\cal H}_{t_n}$ led to
a quantum theory where the notion of time appears
mainly via its partial ordering property
(quasi-temporal behaviour). In particular, there
was no natural way to express the time
translations from one time slot---that refers to
one copy of the Hilbert space ${\cal H}_t$---to
another one, that refers to another copy ${\cal
H}_{t^{\prime}}$. The situation changed when the
{\em the continuous limit\/} of such tensor
products was introduced: hence forward, time
appears uniformly in a continuous way.

One of the original problems in the development of
the HPO theory was the lack of a clear physical
meaning of some of the quantities involved. The
introduction of the history group by Isham and
Linden \cite{IL95} made a significant step in this
direction in the sense that the spectral
projectors of the history Lie algebra represent
propositions about the appropriate phase space
observables of the system.

In standard canonical quantum theory it has been
argued that the identification of Hilbert space
and observables can come from the study of the
representations of the canonical group: a group of
canonical transformations that acts transitively
on the phase space of the classical system
\cite{I83}. For the case of a particle moving on
the real line $\mathR$, the Hilbert space $\cal H$
of the canonical theory carries a representation
of the Heisenberg-Weyl group with Lie algebra
\begin{equation}
    {[\,}x,\, p\,] = i\hbar.           \label{CCR}
\end{equation}

In the history theory, the tensor product ${\cal
V}_n$ of copies of the standard Hilbert space
carries a unitary representation of the $n$-fold
product group whose generators satisfy
\begin{eqnarray}
    {[\,}x_{t_i},\,x_{t_j}\,]&=& 0                  \label{discreteHWxx} \\
    {[\,}p_{t_i},\,p_{t_j}\,]&=& 0                  \label{discreteHWpp} \\
    {[\,}x_{t_i},\,p_{t_j}\,]&=& i\hbar\delta_{ij}  \label{discreteHWxp}
\end{eqnarray}
where $\{t_1, t_2, \ldots , t_n\}$ is a discrete
subset of the real line containing the instants of
time at which the propositions are asserted. Thus,
the tensor product ${\cal V}_n$ can be viewed as
arising as a representation space for Eqs.\
(\ref{discreteHWxx})--(\ref{discreteHWxp}), and
the tensor products
$\alpha_{t_1}\otimes\alpha_{t_2}\otimes\cdots\otimes\alpha_{t_n}$---that
correspond to sequential histories about the
values of position or momentum (or linear
combinations of them)---are then elements of the
spectral representations of this Lie algebra.

In the case of a continuous time label, it is
natural to postulate a continuous analogue of the
history algebra Eqs.\
(\ref{discreteHWxx})--(\ref{discreteHWxp}) in
which the Kronecker delta on the right hand side
of Eq.\ (\ref{discreteHWxp}) is replaced by a
Dirac delta function. This has the striking
consequence that the history propositions about
the system are intrinsically time-averaged
quantities: this means that the physical
quantities cannot be defined at {\em sharp moments
in time\/}. The continuous-time histories
formalism is the basis of the work that follows.

For purposes of clarity, we have found it useful
to briefly review the continuous-time histories
programme. The main ideas are elaborated in section
II together with an explanation of the details of
the general construction. Particular emphasis is
placed on the two distinct laws of time
transformation that arise in history theories, and
their physical interpretation.

In section III we proceed to examine the classical
histories treatment of parameterised systems. We
show that given a constraint $h = 0$ in the
canonical theory we can write the corresponding
function $H_{\kappa}$ in the history theory. The
symplectic transformations generated by
$H_{\kappa}$ partition the space of histories into
orbits and allow us to define the space of reduced
phase space histories. This has the {\it same
temporal structure\/} as the initial space of
histories and hence a clear ordering structure for
time, even for parameterised systems. We will show
that the solutions to the classical equations of
motion are reparameterisation invariant. As an
example we study the system of two harmonic
oscillators with a constant energy difference.

In section IV we proceed to give the quantisation
algorithm. A Hilbert space is constructed by
seeking a representation of the history group in
which the Liouville and Hamiltonian functions are
represented by self-adjoint operators. The
constraint is implemented by projecting any
operator onto the subspace where the quantum
version of the constraints are satisfied. The
decoherence functional is explicitly constructed
and its invariance under time reparameterisation
established. The example of section III is now
quantised and the algorithm is readily implemented
since the constraint operator has a discrete
spectrum.

Section V gives a preliminary account of the case
of a relativistic particle. The canonical
constraint has a continuous spectrum, and hence
the quantisation algorithm has to be modified for
this case. We discuss briefly the strategy to be
followed in a subsequent paper in which we shall
deal with this larger class of systems.

Finally in section VI we discuss and review our
results, putting emphasis on their relevance for
the study of quantum gravity.

We should remark that while in this paper we focus
primarily on parameterised systems, the algorithm
we present is valid for any system with first
class constraints.

\section{The continuous-time histories programme}
As was mentioned in the Introduction, the history
group was first introduced for discrete-time
histories \cite{IL95}, and was then developed to
the continuous time case by introducing a delta
function in the unequal-time history commutation
relations. As an immediate consequence, an
intriguing feature of the theory appeared: that
all interesting history propositions are about
{\em time-averaged} physical quantities.

\subsection{The History Space}

\paragraph*{The History Group.}
To discuss continuous-time histories we employ the
same approach as the one described in the
Introduction. Thus, motivated by Eqs.\
(\ref{discreteHWxx})--(\ref{discreteHWxp}), we
started with the history-group whose Lie algebra
(History Algebra)
 is \cite{IL95}
\begin{eqnarray}
{[\,}x_{t_1},\,x_{t_2}\,] &=&0 \label{ctsHWxx} \\
{[\,}p_{t_1},\,p_{t_2}\,] &=&0 \label{ctsHWpp} \\
{[\,}x_{t_1},\,p_{t_2}\,]
&=&i\hbar\tau\delta(t_1-t_2)\label{ctsHWxp}
\end{eqnarray}
where $-\infty\leq t_1,\,t_2\leq\infty$ ; the
constant $\tau$ has dimensions of {\em time\/}
\cite{Sav99}. It does not appear in any physical quantities; henceforward we
shall set $\hbar \tau = 1$.
Note that these operators are in the
{\em Schr\"odinger\/} picture.

The choice of the Dirac delta function in the
right hand side of Eq.\ (\ref{ctsHWxp}) is closely
associated with the requirement that time be
treated as a continuous variable. As emphasised
earlier, one consequence is the fact that the
observables cannot be defined at sharp moments of
time but rather as time-averaged quantities.

We note that
 Eqs.\
(\ref{ctsHWxx})--(\ref{ctsHWxp}) are
mathematically the same as the canonical
commutation relations of a quantum {\em field \/}
theory in one space dimension. In particular, in
order that equations of the type Eqs.\
(\ref{ctsHWxx})--(\ref{ctsHWxp}) are
mathematically well-defined they must be smeared
with test functions to give
\begin{eqnarray}
    {[\,}x_f,\,x_g\,]&=&0                   \label{SmearedCtsHWxx}  \\
    {[\,}p_f,\,p_g\,]&=&0                   \label{SmearedCtsHWpp}  \\
    {[\,}x_f,\,p_g\,]&=&i \int_{-\infty}^\infty f(t)g(t)\,dt,
                                            \label{SmearedCtsHWxp}
\end{eqnarray}
where the class of test functions $s$ is a linear
subspace of the space $L^2(\mathR, dt)$ of square
integrable functions on $\mathR$. The physically
appropriate representation of the HA Eqs.\
(\ref{SmearedCtsHWxx})--(\ref{SmearedCtsHWxp}),
---bearing in mind that infinitely many unitarily
inequivalent representations are known to exist in
the analogous case of quantum field theory---can be
uniquely selected by the requirement that the
time-averaged energy exists as a proper
self-adjoint operator \cite{A60,ILSS98}.

\paragraph*{Fock representation.}
The selection of the representation by demanding
the existence of a time-averaged energy can be
explicitly carried out for quadratic systems. Here
we shall examine the example of the
one-dimensional, simple harmonic oscillator with
Hamiltonian
\begin{equation}
    H={p^2\over 2m}+{m\omega^2\over 2}x^2.
\end{equation}
We expect to have a one-parameter family of
operators that represent the energy at time $t$
\begin{equation}
H_t:={p_t^2\over 2m}+{m\omega^2\over 2}x_t^2
\label{Def:Ht}
\end{equation}
More precisely, we are interested in a
time-averaged Hamiltonian operator
 $H_\kappa$, which is
defined heuristically as
$H_\kappa=\int_{-\infty}^{\infty}dt\, \kappa(t)
H_t$.

We shall select a Fock representation for the
history algebra that has the structure of a
continuous tensor product. The Fock space is
defined as ${\mathcal V} =e^{L^2(\mathR,dt)}$,
and the generators of the history group are
\begin{eqnarray}
x_t = (\frac{1}{2 \omega})^{1/2} (b_t +
b^{\dagger}_t) \\ p_t = i (\frac{\omega}{2})^{1/2}
(b^{\dagger}_t - b_t)
\end{eqnarray}
in terms of the annihilation and creation
operators $b_t$ and $b^{\dagger}_t$ on the Fock
space. Then the operator $H_{\kappa}$ can be shown
to exist for all measurable functions $\kappa(t)$,
since the automorphisms
\begin{equation}
e^{iH_{\kappa}s} b_t e^{-iH_{\kappa}s} = e^{-i
\omega \kappa(t)s} b_t							\label{automH}
\end{equation}
are unitarily implementable.

The Fock space contains the unnormalised coherent
states $ |\exp w(\cdot) \rangle_{\mathcal V}$
where $w(\cdot) \in L^2(\mathR,dt)$. If one
compares these to the unnormalised coherent states
$ |\exp w \rangle $ of the single-time Hilbert space
$H_t = e^\mathC$ on which the canonical group is
represented, we get the fundamental automorphism
\begin{eqnarray}
\otimes_t H_t &=& {\mathcal V} \nonumber \\
\otimes_t | \exp w_t \rangle_{H_t} &=& | \exp
w(\cdot) \rangle_{{\mathcal V}}.
\end{eqnarray}

\paragraph{The Action and the Liouville operators.}
The question how the Schr\"odinger-picture objects
with different time labels---like the history
algebra generators $x_t$ and $p_t$---are related
was addressed in \cite{Sav99} by studying the time
transformation laws of the theory. The fact that
the Hamilton-Jacobi functional $S$, evaluated on
the realised path of the system---{\em i.e.}, for
a solution of the classical equations of motion,
under some initial conditions---is the generating
function of a canonical transformation which
transforms the system variables, led to the
definition of the action operator
\begin{equation}
 S_{\kappa}:= \int^{+\infty}_{-\infty}( p_{t}\dot{x}_{t}- \kappa(t)H_{t})dt
        \label{Def:op_S}          
  \end{equation}
which can be defined rigorously through its
automorphisms \cite{Sav99}
\begin{equation}
 e^{is S_{\kappa}} b_{t} e^{-is S_{\kappa}}=
e^{i\omega \int^{t+s}_{t} \kappa(t+s^{\prime})
ds^{\prime}}
  e^{s\frac{d}{dt}}b_{t}  \label{Sauto}.
\end{equation}

The first term of the action operator eq.\
(\ref{Def:op_S}) is identical to the kinematical
part of the classical phase space action functional. This  `Liouville'
operator is formally
written as
\begin{equation}
 V:= \int^{\infty}_{-\infty}(p_{t}\dot{x_{t}}) dt \label{liou}
\end{equation}
so that 
\begin{equation}
 S_{\kappa} = V - H_{\kappa}
\end{equation}
The Liouville operator  is rigorously defined through its automorphisms
\begin{equation}
e^{isV}b_te^{-isV} = b_{t+s}.
\end{equation}

\paragraph{General systems.}
This construction can be generalised for more
general systems with quadratic Hamiltonians.
 If the
canonical commutation relations are smeared by
elements of a real vector space $V$, then the
corresponding Hilbert space of the canonical
theory is the Fock space $H = e^{V_c}$, where $V_c
= V \otimes \mathC$. Unnormalised coherent states
$|\exp w \rangle$ with $w \in V_c$ are then
naturally defined on $H$.

The corresponding Weyl group for histories has as
its space of smearing functions the Hilbert space
$ L^2(\mathR,dt)_{\mathR} \otimes V$ and can be
represented in the history Fock space ${\mathcal
V} = e^{L^2(\mathR,dt) \otimes V_c}$. This Fock
space contains the natural unnormalised coherent
states $|\exp w(\cdot) \rangle$ with $w(\cdot) \in
L^2(\mathR,dt) \otimes V_c$ and there is an
isomorphism
\begin{eqnarray}
\otimes_t (e^{V_c})_t &=& e^{L^2(\mathR,dt)
\otimes V_c} \nonumber \\ \otimes_{t}|\exp w_t \rangle_{H_t}
&=& | \exp w(\cdot) \rangle_{{\mathcal V}}
\end{eqnarray}

 For more general Hamiltonians, we can no longer rely on the Fock
type of representations. The selection of a
representation is particularly difficult when the
Hamiltonian has only continuous spectrum. This
will be a problem in the quantisation of the
relativistic particle.

\subsection{The temporal structure}

We have seen that the Liouville and the Hamilton
operators generate two distinct laws of time
transformation. This is a fundamental property of
history theories. In this section we shall explain
their physical meaning \cite{Sav99,Sav99b}.

In the continuous-time histories theory, the
Hamiltonian operator $H_t$ produces phase changes
in time, preserving the time label $t$ of the
Hilbert space on which, at least formally, $H_t$
is defined. On the other hand, and analogous to
the classical case, it is the Liouville operator
$V$ that assigns history commutation relations,
and produces time transformations `from one
Hilbert space to another'. The action operator
generates a combination of these two types of time
transformation.

This is easier to see through the definition of a
Heisenberg-picture analogue of $x_{t}$
\begin{eqnarray}
x_{\kappa,t,s } :&=& e^{i s
H_{\kappa}} x_{t} e^{-isH_{\kappa}}
\\ \label{heis}
&=& \cos[\omega s \kappa(t)]x_{t} + \frac{1}{m\omega}\sin[\omega s
\kappa(t)]p_{t}
\end{eqnarray}
and the similar definition for $p_{\kappa,t,s}$.

If we use the notation $x_{f}(s)$ for the {\em
history} Heisenberg-picture operators smeared with
respect to the time label $t$ Eq.\ (\ref{heis})
(with $\kappa=1$), we observe that they behave as
standard Heisenberg-picture operators, with time
parameter $s$.

We now define a one-parameter group of
transformations $T_{V}(\tau)$, with elements
$e^{i\tau V}$, $\tau\in{\mathbf{R}}$
where $V$ is the Liouville operator Eq.\
(\ref{liou}), and we consider its action on the
Heisenberg-picture operator $b_{t,s}$ (for
simplicity we write the unsmeared expressions)
\begin{equation}
 e^{i \tau V} b_{t,s} e^{ - i \tau V} =
 b_{t+\tau,s},
\end{equation}
which makes particularly clear the sense in which
the Liouville operator is the generator of
transformations of the time parameter $t$ labelling
the Hilbert spaces ${\mathcal{H}}_{t}$.

Next we define a one-parameter group of
transformations $T_{H}(\tau)$, with elements
$e^{i\tau H}$, where $H$ is the
 Hamiltonian operator smeared with the function $\kappa(t) = 1$
\begin{equation}
 e^{i\tau H} b_{t,s} e^{ - i \tau H} =
 b_{t,s+\tau}.
\end{equation}
Thus the Hamiltonian operator is the generator of
phase changes of the time parameter $s$, produced
only on one Hilbert space ${\mathcal{H}}_{t}$, for
a fixed value of the `external' time parameter
$t$.

 Finally, we define the one-parameter group of
transformations $T_{S}(\tau)$, with elements
$e^{i\tau S}$, where $S$ is the
action operator, which acts as
\begin{equation}
 e^{i\tau S} b_{t,s} e^{ - i\tau S} = b_{t+\tau ,
 s+\tau}
\end{equation}
We see that the action operator generates both
types of time transformations---a feature that
appears only in the HPO scheme.

In standard quantum theory, time evolution is
described by two different laws: the state-vector
reduction that supposedly occurs when a
measurement is made, and the unitary
time-evolution that takes place between
measurements.

 It is our contention that the two types of
time-transformation observed in the
continuous-time histories are associated with the
two dynamical processes in standard quantum
theory: the time transformations generated by the
Liouville operator $V$ are related to the causal
ordering implied by the temporal logic nature of
the histories construction (we shall argue in
section II.4 that it is related to the
state-vector reduction in the canonical theory),
while the time transformations produced by the
Hamiltonian operator $H$ are related to the
unitary time-evolution between ``measurements''.

The Hamiltonian operator, which produces
transformations via a type of Heisenberg
time-evolution, appears as the `clock' of the
theory. As such, it depends on the particular
physical system that the Hamiltonian describes.
Indeed, we would expect the definition of a
`clock' for the evolution in time of a physical
system to be connected with the dynamics of the
system concerned.

We note that the smearing function $\kappa (t)$
used in the definition of the Hamiltonian operator
can be interpreted as a mechanism for implementing
the idea of reparameterising time; in the present
context however, $\kappa$ is kept fixed for a
particular physical system.

The coexistence of the two types of
time-evolution, as reflected in the action
operator identified as the generator of such time
transformations, is a striking result. In
particular, its definition is in accord with the
classical analogue, namely the Hamilton action
functional. In classical theory, a distinction
also arise between a kinematical and a dynamical
part of the action functional in the sense that
the first part corresponds to the symplectic
structure and the second to the Hamiltonian.

\subsection{The classical histories}

Let us consider the space of classical histories
$\Pi$ viewed as the set of smooth\footnote{The
requirement of smoothness is made for simplicity,
but it might be appropriate to extend the paths to
a larger class of function for certain purposes;
for example, in a careful discussion of the
temporal logic encoded in the classical theory.} paths on the classical phase space
$\Gamma$. Hence an element of $\Pi$ is a path
$\gamma:\mathR \rightarrow \Gamma$. For our
purposes we shall have to define vector fields and
differential forms on $\Pi$. Therefore one has to
equip $\Pi$ with the structure of tangent and
cotangent spaces at one point. To avoid the need
to introduce a specific (infinite-dimensional)
differential structure on $\Pi$, we use the
standard trick of employing the following
definitions. The tangent space $T_{\gamma} \Pi $
at $\gamma \in \Pi$ is defined as
\begin{equation}
T_{\gamma} \Pi = \{ v : \mathR \rightarrow
T\Gamma\mid v(t) \in T_{\gamma(t)} \Gamma \}    \label{tangent}
\end{equation}
where $T \Gamma$ is the tangent bundle of
$\Gamma$. Correspondingly the cotangent space at
$\gamma$ is defined as
\begin{eqnarray}
T^*_{\gamma} \Pi = \{ \omega: \mathR \rightarrow
T^* \Gamma \mid\omega(t) \in T^*_{\gamma(t)}
\Gamma\}											\label{cotangent}
\end{eqnarray}
and the pairing between covariant and
contravariant vectors at the point $\gamma\in\Pi$
is given by
\begin{equation}
\langle \omega(\cdot), v(\cdot) \rangle_{\gamma} =
\int dt \langle \omega(t), v(t)
\rangle_{\gamma(t)}
\end{equation}
where the brackets within the integral denote the
standard pairing on $\Gamma$ (thus the functions
$v(\cdot)$ and $\omega(\cdot)$ must be such that
this integral exists).

Taking for simplicity $\Gamma = \mathR \times
\mathR = \{ (x,p) \}$, we can define $x_t$ and
$p_t$ as functions on $\Pi$ by
\begin{eqnarray}
x_t (\gamma) &:=& x(\gamma(t)) \\ p_t(\gamma)
&:=&p(\gamma(t))
\end{eqnarray}
In general given a function $f$ on $\Gamma$ we can
define a family $t\mapsto F_t$ of functions on
$\Pi$ as
\begin{equation}
F_t(\gamma) := f(\gamma(t))            \label{Ff}
\end{equation}
We can equip $\Pi$ with a symplectic form
\begin{equation}
\omega = \int dt dp_t \wedge dx_t
\end{equation}
which generates a Poisson bracket
\begin{equation}
\{ x_t,p_{t'} \}_{\Pi} = \delta(t,t').
\end{equation}
More generally, for two families of functions
$t\mapsto F_t$ and $\mapsto G_t$ defined through
(\ref{Ff}) we have
\begin{equation}
\{F_t,G_{t'} \} = L_t \delta(t,t').
\end{equation}
where $L_t$ corresponds to the function $l$ on
$\Gamma$
\begin{equation}
l = \{ f,g \}_{\Gamma}.
\end{equation}
We now define the phase space action functional
$S(\gamma)$ on
 $\Pi$ as
\begin{equation}
   S(\gamma):=
   \int{[p_t\dot{x_t}-H_t(p_t,x_t)](\gamma)\,dt}
\end{equation}
where $\dot{x_t}(\gamma)$ is the velocity at the
time point $t$ of the path $\gamma$.

We also define the history classical analogues for
the Liouville and time-averaged Hamiltonian
operators as
\begin{eqnarray}
 V ( \gamma )&:=& \int{[p_t\dot{x_t}](\gamma)\,dt}    \\
 H ( \gamma )&:=& \int{[H_t(p_t,x_t)](\gamma)\,dt}    \\
   S( \gamma ) &=& V( \gamma )- H( \gamma )
\end{eqnarray}

In classical mechanics, the least action principle
states that there exists a functional $S(\gamma )=
\int{[p\dot{x}-H(p,x)](\gamma)\,dt}$ such that the
physically realised path is a curve in state
space, ${\gamma}_{0}$, with respect to which the
condition $\delta S ({\gamma}_{0}) = 0$ holds,
when we consider variations around this curve.
From this, the Hamilton equations of motion are
deduced to be
\begin{eqnarray}
  \dot{x} &=& \{ x , H \}         \\
  \dot{p} &=& \{ p , H \}
\end{eqnarray}
where $x$ and $p$---the coordinates of the
realised path ${\gamma}_{0}$---are the solutions
of the classical equations of motion \cite{Sav99, Sav99b}. For any
function $F(x,p)$ of the classical solutions it is
also true that
\begin{equation}
    \{F, H \} = \dot{F}
\end{equation}

In the case of classical continuous-time
histories, one can formulate the above variational
principal in terms of the Hamilton equations with
the statement: A classical history ${\gamma}_{cl}$
is the realised path of the system---{\em i.e.\/}
a solution of the equations of motion of the
system---if it satisfies the equations
\begin{eqnarray}
     \{x_t , V\}(\gamma_{cl}) = \{x_t , H\}(\gamma_{cl}) \label{Ham1}
     \\
      \{p_t , V\}(\gamma_{cl}) = \{p_t , H\}(\gamma_{cl})
      \label{Ham2}
\end{eqnarray}
where ${\gamma}_{cl}= t\mapsto(x_t({\gamma}_{cl})
, p_t({\gamma}_{cl}) )$, and $x_t({\gamma}_{cl})$
is the position coordinate of the realised path
${\gamma}_{cl}$ at the time point $t$. The eqs.\
(\ref{Ham1}--\ref{Ham2}) are the history
equivalent of the Hamilton equations of motion.

One would have expected this result for the
classical analogue of the histories formalism, as
it shows that the classical analogue of the two
types of time-transformation in the quantum theory
coincide.

 From the eqs.\
(\ref{Ham1}--\ref{Ham2}) we also conclude that the
canonical transformation generated by the history
action functional $S(\gamma)$, leaves invariant
the paths that are classical solutions of the
system:
\begin{eqnarray}
     \{x_t , S \}(\gamma_{cl})&=& 0  \label{clsol1} \\
     \{p_t , S \}(\gamma_{cl})&=& 0  \label{clsol2}
\end{eqnarray}
It is also the case that any function $F$ on $\Pi$
satisfies the equation
\begin{equation}
     \{ F, S\} (\gamma_{cl})= 0.
\end{equation}

\subsection{The general form of the decoherence functional}
Having identified the distinct notions of time
structure in histories theory, we now show how
they are manifested in the assignment of
probabilities. In the case of discrete time, given
the histories Hilbert space ${\mathcal V}$ one can
define the boundary Hilbert space $\partial
{\mathcal V}$ consisting of bounded maps from the
initial time Hilbert space $H_{t_i}$ to the final
time Hilbert space $H_{t_f}$. Then the decoherence
functional can be written as \cite{An00}
\begin{equation}
d(\alpha, \alpha') = Tr_{\partial {\mathcal V}}
\left( c(\alpha) c^{\dagger}(\alpha') \right)
\end{equation}
where $c(\alpha)$ is an operator on $\partial
{\mathcal V}$ defined by
\begin{equation}
c(\alpha) = Tr_{{\mathcal V}} \left( {\cal AS}
{\cal U}^{\dagger} \alpha {\cal U} \right)                   \label{decfun}
\end{equation}
${\cal S}$ and ${\cal U}$ are operators on
${\mathcal V}$ containing the kinematical/causal
and dynamical structure respectively. They are
defined in the discrete-time case as
\begin{eqnarray}
{\cal S} |v_{t_1} \rangle | v_{t_2} \rangle \ldots
| v_{t_n} \rangle &=& |v_{t_n} \rangle | v_{t_1} \label{Schr_op}
\rangle \ldots | v_{t_{n-1}} \rangle \\ 
{\cal U} &=&
U(t_1) \otimes U(t_2)\otimes \cdots \otimes U(t_n)   \label{Heisen_op}
\end{eqnarray}

The physical meaning of these operators in the
decoherence functional also highlights the
temporal structure of history theories. The operator ${\cal S}$ incorporates the contribution
of the system's dynamics to the assignment of
probabilities. Its action is to transform a
Schr\"odinger picture operator to a Heisenberg
picture one, and it does not change the time label
$t$. The operator $\cal S$ plays essentially the
role of a connection, identifying the structures
of Hilbert space at successive instants of time.
It is therefore closely related to the Liouville
operator. In fact, ${\cal S }$ plays a role which
is associated with ``wave-packet reduction'' in
standard canonical quantum theory: it essentially
forces the single-time projectors that form a
history to be multiplied (when forming the class
operator $\tilde{C}$) in the order provided by the
partial ordering of time.

The operator ${\cal A}$ is a
 linear map from ${\mathcal V}$ to
$\partial {\mathcal V}$. It incorporates the
effects of initial conditions (or, if appropriate,
final ones also). In the case of discrete time, if
 $\{|v_{t_1} \ldots v_{t_n} \rangle\}$ denotes an orthonormal
basis of the history Hilbert space, the matrix
elements of ${\cal A} $ are given by
\begin{eqnarray}
\langle v'_{t_1} \ldots v'_{t_n} |{\cal A}^{ij}|
v_{t_1} \ldots v_{t_n} \rangle =
\frac{(\rho_f^{1/2})_{v'_{t_1}i} (\rho_0^{1/2})_{j
v_{t_1}}}{\tr(\rho_f \rho_i)} \delta_{v'_{t_2}
v_{t_2}} \ldots \delta_{v'_{t_n} v_{t_n}}   \label{Bound}
\end{eqnarray}
where the indices $i,j$ label the orthonormal
basis $\{|v_{t_1} \rangle \otimes |v_{t_n}
\rangle\}$ on the boundary Hilbert space.

The above expressions are valid for the
discrete-time case as can be checked by direct
substitution and   with
Eq. (\ref{decfun1}). With
care, they can be generalised to the case of
continuous time. The operator ${\cal S}$ can be
constructed from Eq. (\ref{Schr_op}) by going to the continuum
limit. The operator ${\cal U}$ cannot readily be
defined as a continuous-time tensor product of
constituent operators; however, Eq. (\ref{Heisen_op}) suggests
that it should be identified with $e^{-iH_{id}}$,
where $id(t) = t$. And, in the case of the
operator ${\cal A}$, one can keep a definition
similar to (\ref{Bound}) with a particular choice of
coherent-state basis \footnote{We have to take
into account the fact that, for continuous time,
the boundary Hilbert space cannot strictly
speaking be embedded into ${\cal V}$, as is
possible in the case of discrete time.
Nonetheless, it can still be considered as
consisting of two copies of the single-time
Hilbert spaces. This is the most conservative
position, but it is important to note that in the
presence of continuous time the initial state
might be encoded in more general types of object
than density matrices defined at sharp moments of
time.} (for an example see reference \cite{An00}).

\section{Parameterised systems: the classical case}

\subsection{Canonical treatment}

The starting point for the Hamiltonian treatment
of a constrained system is the phase space
$\Gamma$ of the unconstrained system. A key
feature is the existence of a first-class
constraint $h(x,p) = \xi$, where $\xi$ is a
constant \footnote{We make the choice of $\xi$ so
that the function $h$ does not contain any
constant terms. This will be convenient in the
histories treatment.}. By this we mean that the
physical system is constrained to lie in the
appropriate subset of $\Gamma$: namely, the
constraint surface $C$. What is special in a
parameterised system is the fact that the
Hamiltonian is itself a multiple of $h(x,p) -
\xi$. Hence it vanishes on the constraint surface.

The fact that the constraint is first class means
that the function $h$ is a generator of a
`symmetry' of the system. By this we mean that two
points of the constrained surface that are related
by a canonical transformation generated by $h$
correspond to the same physical state of the
system. In this context we note that the
constraint surface is not itself a symplectic
manifold since the restriction of the symplectic
form to $C$, $\omega_C$, is degenerate. This means
that the points in $C$ are not in one-to-one
correspondence with the physical states of the
system. The true degrees of freedom lie in the
{\em reduced phase space}. This is constructed as
follows.

We first identify the orbits of the one-parameter
group of canonical transformations generated by
the constraint. The set of all such orbits on $C$
is the reduced phase space $ \Gamma_{red}$. We
denote by $c_p(s)$ the curve obtained when the
one-parameter group of canonical transformations
generated by $h$ acts on $p \in \Gamma$. Clearly
$c_p(0) = p$. Then we define the equivalence
relation $\sim$ as follows
\begin{equation}
\mbox{$p \sim p'$ for $p, p' \in \Gamma$ if there
exists $s \in \mathR$ such that $c_p(s) = p'$.}
\end{equation}

This relation partitions $\Gamma$ into orbits. The
fact that $h$ is a first-class constraint implies
that all orbits with a point in the constraint
surface lie wholly in the constraint surface. We
can therefore define the reduced phase space as
$\Gamma_{red} = C/ \sim$.
 The symplectic
form that is naturally defined on $\Gamma_{red}$
is non-degenerate, since the vector field that
generates the action of the constraints lies in
the degenerate direction of $\omega_C$.

\subsection{The classical problem of time}
The above prescription for the construction of the
reduced phase space holds for a general
constrained system. But when the system is
parameterised there is a conceptual difficulty
concerning the description of time evolution in
$\Gamma_{red}$. First, the elements of reduced
phase space are themselves solutions to Hamilton's
equation, since the Hamiltonian is proportional to
the constraint. On the other hand, a point of the
reduced phase space ought to correspond to a state
of a system at an instant of time. This is not the
case for a parameterised system. Indeed, a point
of the reduced phase space is a whole {\em
history\/} and is not restricted to a single
moment of time. This means that we do not know
what time evolution means in these systems, {\em
i.e.}, what we mean when we say that a point of
$\Gamma_{red}$ `evolves' in time.

Parameterised systems are therefore
 intrinsically timeless: this is not only because their  Hamiltonian
vanishes---this would just imply that the dynamics
is trivial. But even at the {\em kinematical\/}
level of the definition of what is meant by time
evolution there exists an ambiguity. We cannot
think of the basic observables of a parameterised
system as quantities that have even the {\em
potentiality\/} of evolving.

This is the classical `problem of time'. One way
of tackling this is to impose a gauge-breaking
condition that selects a time variable. This
essentially amounts to selecting a degree of
freedom as the `clock' of the theory. The most
elegant implementation of this idea is due to
Rovelli \cite{Rov90,Rov91}. A gauge is chosen in such a
way that a function on the constraint surface
(that does not commute with the constraint) is
identified as a physical clock of the system. One
then can assign a time parameter as the value of
the associated observable.

There are many problems with this line of thought.
First, the choice of the clock variable involves
an arbitrary choice of gauge that cannot be
justified {\em a priori}, and when going to the
quantum version of this system there is no
guarantee that the physical results will be
independent of the choice. Secondly, the topology
of the constraint surface might not allow us to
find a variable that can play the role of a good
clock: {\em i.e.}, one that does not repeat the
same values of time in the course of evolution.
For example, this is the case when the constraint
surface is compact \cite{Rov90} since it cannot
have the whole real line $\mathR$ as a
submanifold.

Also, Hartle has severely criticised the notion
that clock time is fundamental \cite{Har96}. His
argument is mainly quantum mechanical: that
arbitrary choices of time function cannot
reproduce the results of standard quantum theory
since they might put possible events in a {\em
different\/} causal relation. Overall, it seems
that a `clock' time---all that is available in the
canonical description of a parameterised
system---might provide a numerical measure of
change, but it is unable to incorporate the causal
aspect of time in which it is expected to be a
partial ordering that determines the succession of
events.

From the perspective of the histories programme,
the origin of this problem is clear. The canonical
approach conflates the two conceptually distinct
aspects of time, within the guise of Hamiltonian
time evolution. Once they are distinguished the
problem is immediately resolved. Histories are
paths, and observables are labelled by the
external variable $t$, which contains the notion
of partial ordering. When the constraints are
implemented, the reduced objects are still
histories labelled by the same parameter. In
parameterised systems it transpires that the
precise value of $t$ is not important: all that
matters is that the causal ordering is preserved.
This means that the theory is invariant under
reparameterisations of time.

We can still define clocks (or construct clocks in
practice) as we do in classical mechanics by
picking some observable whose values label the
changes undergone by the system. But clock time
does not exhaust the properties of time---it is
just a quantification of change. Of course, the
value that specifies the duration of a given
process is not arbitrary but depends on the
structure of the physical system (including the
clock) we are studying: {\em i.e.}, its
Hamiltonian.

The histories perspective towards time is similar
(but not identical) to the Aristotelian
distinction between temporality as change and
process ({\em kinesis, physis}) and time as
measure of this process ({\em chronos}).

\subsection{Histories perspective}
\subsubsection{The space of paths}
We shall now describe the parameterised system in
the histories language. As discussed in Section
II, the fundamental object is the space of
classical histories $\Pi$, consisting of paths
$\gamma: \mathR \rightarrow \Gamma$. This can be
equipped with a symplectic form $\Omega$ which can
be written in terms of the canonical coordinates
$x^i_t $ and $p^j_t$ as
\begin{equation}
\Omega = \int dt \sum_i dp^i_t \wedge d x^i_t
\end{equation}
Equivalently, the fundamental Poisson bracket on
$\Pi $ is
\begin{equation}
\{x^i_t, p^j_{t'} \} = \delta^{ij} \delta(t,t')
\end{equation}
The functions $x_t^i$ are a special case of a
construction of families of observables labelled
by $t$. That is, given a function $f$ on $\Gamma$
one defines a family of functions $F_t$ on $\Pi$
as
\begin{eqnarray}
F_t(\gamma) := f(\gamma(t))
\end{eqnarray}

\subsubsection{Functions on $\Pi$}
Any function $H$ on $\Pi$ generates a
one-parameter group of canonical transformations.
We shall denote this as $s\mapsto T_H(s)$, and its
action by automorphisms on the algebra of
functions $F$ on $\Pi$ as
\begin{equation}
    F \rightarrow T_H(s)[F].
\end{equation}
\par
In particular, the functions that generating time
translations---$V$, $H_{\kappa}$ and
$S_{\kappa}$---are naturally defined on $\Pi$ as
\begin{eqnarray}
V &=& \sum_i \int dt p_{it} \dot{x}_{it} \\
H_{\kappa} &=& \int dt \kappa(t) h(x_t,p_t) \\ 
S &=& V - H_{\kappa}
\end{eqnarray}

Now $H_{\kappa}$ is the generator of the symmetry
of the system. As $\kappa(t)$ varies, $H_{\kappa}$
form the generators of an infinite-dimensional
Abelian group that acts on $\Pi$ by the canonical
transformations $T_{H_{\kappa}}(s)$.

\subsubsection{The reduction procedure}
We now restrict ourselves to the paths on the
constraint surface by imposing $h(x_t,p_t)(\gamma)
= \xi $ for all $t$.
 In the special case
that $\xi = 0$ this is equivalent to the condition
that $H_{\kappa} = 0$ for all $\kappa$, since the
set of all measurable functions separates
$\mathR$. By this means, the history constraint
surface $C_{h}$ is defined as the space of maps
from $\mathR$ to $C$.

The next step is to study the orbits of the group
of constraints on $\Pi$. We define the equivalence
relation $\sim$ in $\Pi$ as follows:
\begin{eqnarray}
\mbox{$\gamma_1 \sim \gamma_2$ if there exists a
measurable function $\kappa$ such that $\gamma_1 =
T_{H_{\kappa}}(s) \gamma_2$}. \nonumber
\end{eqnarray}
Then the quotient $C_h/\sim := \Pi_{red}$ is the
space of equivalence classes of paths paths. Its
elements can be written as $\tilde{\gamma} =
[\gamma]$, {\em i.e.}, the equivalence class which
contains $\gamma$.

The transformations $T_{H_{\kappa}}(s)$ generated
by $H_{\kappa}$ preserve the $t$ label in a path,
since for any family $\{F_t\mid t\in\mathR\}$
defined by Eq. (\ref{Ff}) we
have
\begin{equation}
    \{H_{\kappa},F_t \} = \kappa (t)G_t.
\end{equation}
The function $G_t$ is defined by (II.3) in terms
of the function $g \in C^{\infty}(\Gamma)$ given
by
\begin{eqnarray}
g = \{h,f \}.
\end{eqnarray}
Therefore $\Pi_{red}$ is identical to the space of
paths on the reduced phase space, {\em i.e.}, maps
$\mathR \rightarrow \Gamma_{red}$. It also
inherits a natural symplectic structure
$\Omega_{red}$. It is a standard theorem of
symplectic reduction that for any function $F$ on
$\Pi$ that commutes weakly with $H_{\kappa}$
\footnote{This means that the commutator
$\{H_{\kappa},F\}$ vanishes on the constraint
surface.} there corresponds a unique function
$\tilde{F}$ on $\Pi_{red}$. This is constructed as
follows. If $F$ commutes with $H_{\kappa}$ it is
constant on its orbits; hence if we denote an
orbit as $[\gamma]$, we can define
\begin{equation}
\tilde{F}([\gamma]) = F(\gamma)
\end{equation}
It is a standard result that the reduction
preserves the Poisson brackets:
\begin{equation}
\widetilde{ \{F,G \} } = \{ \tilde{F}, \tilde{G}
\}.
\end{equation}
In particular, this holds for the families
$\{F_t\mid t\in\mathR\}$ defined earlier.

\subsubsection{The action principle}
In the absence of constraints the paths that
satisfy the equations of motion are obtained by
the requirement
\begin{equation}
\{S, F \} (\gamma) = 0
\end{equation}
for all $F$.
\par
A similar condition could be imposed in the
present case. But it is necessary to reduce the
action to a function on $\Pi_{red}$. This is
possible only if it is compatible with the
equivalence relation $\sim$ generated by the group
of constraints.
 This implies that
\begin{equation}
[ T_{S_{\kappa}}(s)(\gamma)] = [T_{S_{\kappa}}(s)(\gamma')]
\end{equation}
if $\gamma \sim \gamma'$. Therefore, there should
exist a function $\lambda'(\cdot)$ such that
\begin{equation}
T_{S_{\kappa}}(s) T_{H_{\lambda}}(s') = T
_{H_{\lambda'}}(s') T_{S_{\kappa}}(s)    \label{condition}
\end{equation}
for all functions $\kappa(\cdot)$,
$\lambda(\cdot)$ and $s,s' \in \mathR$.

We have that $\{ S_{\kappa}, H_{\lambda} \} =
H_{\dot{\lambda}} $, which indeed implies that by
defining $\lambda' $ as $\lambda'(t) =
\lambda(t+s)$, equation (\ref{condition}) is
satisfied. Hence $S_{\kappa}$ can be projected to
a function $\tilde{S}$ on $\Pi_{red}$ which will
determine the classical paths as the elements
$\gamma$ of $\Pi_{red}$ that satisfy
\begin{equation}
\{ \tilde{S}, F \} (\gamma) = 0
\end{equation}
for all functions $F$.

In fact, $\tilde{S}$ contains only the Liouville
part since it transforms all families
$\tilde{F}_t$ defined by (\ref{Ff}) to
$\tilde{F}_{t+s}$. This comes from the fact that
\begin{eqnarray}
\{ \tilde{S},\tilde{F}_t \} = \widetilde{\{S,F_t
\}} = = \widetilde{\{ V - H_{\kappa}, F_t \}} =
\widetilde{\{V, F_t\}} = \{
\widetilde{V},\tilde{F}_t \}.
\end{eqnarray}

\subsubsection{ Reparameterisation invariance}
The Liouville operator on the reduced phase space
determines the classical solutions. These are
characterised by reparameterisation invariance,
whose generator is a smeared version of the
Liouville operator
\begin{eqnarray}
V_{\lambda} = \int dt\,\lambda(t) p_t \dot{x}_t.
\end{eqnarray}
This satisfies
\begin{equation}
\{V_{\lambda}, H_{\kappa} \} = H_{\dot{\lambda}
\kappa + \dot{\kappa} \lambda}
\end{equation}
and therefore can be projected to a function
$\tilde{V}_{\lambda}$ on $\Pi_{red}$. Clearly we
have
\begin{equation}
\{\tilde{V}, \tilde{V}_{\lambda} \} =
V_{\dot{\lambda}}
\end{equation}

The action of $V_{\lambda}$ is as follows. If we
define
\begin{equation}
\tau_\lambda(t):= \int_a^t \frac{ds}{\lambda(s)}
\end{equation}
($a$ is an arbitrary number) we can reparameterise
the paths from the variable $t$ to the variable
$\tau$ as long as $\tau_\lambda(t) $ is a strictly
increasing function of $t$. We can then take the
inverse $t(\tau)$. This means we write
$\gamma'(\tau) = \gamma(t(\tau))$.
 Then we can write the family of functions
\begin{equation}
F_{\tau} := f(\gamma'(\tau)) = f(\gamma(t(\tau)))
\end{equation}
in analogy to Eq. (\ref{Ff}). Clearly then
\begin{equation}
V_{\lambda} = \int d \tau\, p_{\tau} \frac{d
x_{\tau}}{d \tau}
\end{equation}
 The action of the one-parameter group of transformations $V_{\lambda}$ is
 then
\begin{equation}
F_{\tau} \rightarrow F_{\tau + s}.
\end{equation}
This means that $V_{\lambda}$ generates
translations in the reparameterised time variable
$\tau$. Or equivalently
\begin{equation}
F_t \rightarrow F_{ \tau^{-1}_{\lambda} (
\tau_{\lambda}(t) + s) }
\end{equation}
This equation implies that for the classical paths
({\em i.e.}, those invariant under the action of
the Liouville function) are left invariant under
the action of the smeared Liouville function.
Hence reparametrisations of time leave invariant
the solutions to the classical equations of
motion.

In order for $\tau_{\lambda}$ to be strictly
increasing, its derivative has to be everywhere
positive and non-zero; hence $\lambda(t) $ has to
be positive and non-zero. Then, for each
$\lambda(\cdot)$, $V_{\lambda}$ generate a
semigroup \footnote{It lacks inverses because
$\lambda(t)$ must be positive} of canonical
transformations on $\Pi$, with a corresponding
$\tilde{V}_{\lambda}$ on $\Pi_{red}$ which
generate translations in reparameterised time
under which the classical solutions are invariant.

 \subsection{Harmonic oscillators with constant energy difference}
We shall now give an example of the construction
described above. It is a parameterised system that
has been extensively studied,
 because its quantisation is particularly simple.
This consists of two harmonic oscillators
constrained to have a constant energy difference.

Here the phase space is $\mathR^4$. It is spanned
by the global coordinate
 system $x_0,x_1,p_0,p_1$. The basic Poisson bracket is
 \begin{equation}
 \{ x_i,p_j \} = \delta_{ij}
 \end{equation}
 where $i,j = 0,1$.
We also have the first-class constraint $ h =
\frac{1}{2} (p_0^2 + x_0^2 - p_1^2 - x_1^2) = E$.
The Hamiltonian of the system is proportional to
$h - E$.

 It is more convenient to use complex coordinates $w_0 = (x_0+ i
 p_0)/\sqrt{2}$ and
 $w_1 = (x_1 + i p_1)/ \sqrt{2}$ so that
 \begin{equation}
 h = w^*_0 w_0 - w^*_1 w_1 = E.
 \end{equation}
 \par
 In the history
 version, a path $\gamma \in \Pi$ is parameterised by the coordinates
 $(x_{0t},p_{0t},x_{1t}, p_{1t})$. The space $\Pi$ carries  a symplectic form
 \begin{equation}
 \Omega = \int dt \sum_{i = 0}^1 dp_{it} \wedge dx_{it}
 \end{equation}
 The corresponding Poisson bracket is
 \begin{equation}
 \{ x_{it}, p_{jt'} \} = \delta_{ij} \delta(t,t').
 \end{equation}

 We can also coordinatise $\Pi$ with  $(w_{0t},w_{1t})$. In terms of
these, the
 symplectic form reads
 \begin{equation}
 \Omega = i \int dt\, dw^*_{0t} \wedge dw_{0t} + dw^*_{1t} \wedge dw_{1t}
 \end{equation}
 The action of $H_{\lambda}$ on  $\Pi$ is given by
 \begin{equation}
 (w_{0t}, w_{1t}) \rightarrow ( w_{0t}e^{-i\lambda(t)} , w_{1t} e^{i
 \lambda(t)} )
 \end{equation}
 We use  coordinates $a_t, \phi_t, \chi_t$ on the constraint surface that are
 defined by
 \begin{eqnarray}
 w_{0t} := \sqrt{E} \cosh a_t e^{i \phi_t} \\
 w_{1t} := \sqrt{E} \sinh a_t e^{ i \chi_t}
 \end{eqnarray}
 The degenerate two-form $\Omega_C$ then reads
 \begin{equation}
 \Omega_C  =   - E \int dt\, \sinh 2a_t ( da_t \wedge d \phi_t + da_t \wedge
 d
 \chi_t )
 \end{equation}
 The parameters $\psi_t := (\phi_t + \chi_t)/2$ and $a_t$
 are constant on the $H_{\kappa}$
 orbits, hence they can be used as coordinates on the reduced manifold.
 The coordinates $ \zeta_t := (\phi_t - \chi_t)/2$ correspond to the
 degenerate
 directions of $\Omega_C$. We can therefore parameterise $\Pi_{red}$ by
 $\psi_t$ and $ c_t = E \cosh 2 a_t $, and write the symplectic form
 $\Omega_{red}$ in
 the local coordinate form
 \begin{equation}
 \Omega_{red} = \int dt\, d \psi_t \wedge dc_t.
 \end{equation}
 The action $S_{\kappa}$ acts on $\Pi$ as
 \begin{equation}
 (w_{0t}, w_{1t}) \rightarrow  (w_{0t+s} \exp ( -i \int_0^s \kappa(t+s')ds'),
 w_{1t+s}
 \exp ( i \int_0^s \kappa(t+s') ds') )
 \end{equation}
 which on $\Pi_{red}$ projects as
 \begin{equation}
 ( \psi_t, c_t) \rightarrow (\psi_{t+s}, c_{t+s})
 \end{equation}
 which identifies the elements $\tilde{\gamma}$ of $\Pi_{red}$ as the ones
 satisfying
 \begin{eqnarray}
 \psi_t (\tilde{\gamma}) = const. \\
 c_t (\tilde{\gamma}) = const.
 \end{eqnarray}
Clearly the Liouville function on $\Pi_{red}$
is equal to
\begin{equation}
\tilde{V} = \int dt\, c_t \dot{\psi}_t.
\end{equation}

 \section{The quantum treatment}
  \subsection{Reduced phase space quantisation}
   The logic of reduced phase space quantisation is to
focus on the reduced phase space and forget the
classical procedure by which one arrived there.
Then one can use some quantisation algorithm to
construct a quantum theory, that will contain
observables corresponding to ones defined on the
reduced phase space. In this approach one has to
identify
   the temporal structure already at the classical level (for example,
through a time
   function as explained in section III.2). This means we would
   inherit the problems associated with lack of gauge invariance and loss of temporal ordering that characterise the classical problem of time.

 In the histories quantisation one should first look for a a history group on $\Pi_{red}$.
   This  will be in general different from the Weyl group, and its structure will
   depend on the topology of the reduced phase space. Then we should
   construct
the history Hilbert space, by requiring that it
carries a representation of the history group.

In general $\Pi_{red}$ has a non-trivial topology.
This implies that the choice of the history group
is complicated, and for this reason we have chosen
to work with a Dirac quantisation scheme.

   \subsection{Dirac quantisation}
   In the canonical quantisation scheme the Dirac method consists in
   finding the Hilbert space that carries  a unitary representation of
the appropriate
   canonical group. We further require that the  classical constraint is
   represented
   by a self-adjoint operator  $h$. Then the physical Hilbert space is the
   linear
   subspace of the Hilbert space that corresponds to the $\xi$-eigenvalue of
   the
   constraint operator $h$.

   \subsubsection{The quantum mechanical problem of time}
   Even if we ignore the technical problems of  the case where the
   constraint has a continuous spectrum,
   the Dirac formalism in the canonical setting suffers from
   the
   problem of time.
   As emphasised earlier, there is no natural notion of time evolution in the
   physical Hilbert
   space, since the Hamiltonian  vanishes there. As such we cannot speak
   about time evolution either of states or of observables in the physical
   Hilbert space. A possibility that has been proposed in the canonical
   context is that time might be identified with some quantum mechanical
   observable. This is again a notion of `clock time'. This is even more
   difficult
   to accept in quantum theory, because any physical clock inevitably
undergoes
   quantum fluctuations, and there is no guarantee that it would be able to
   respect the temporal ordering of events. To this, one
   can add the fact that the quantum theory would be explicitly non
   gauge-invariant.

One proposal that partially remedies this problem,
is that clock time is meaningful only in a
classical limit; for instance when the wave
function describing the system is of WKB type.
(This is very prominent in the discussions of the
Wheeler-DeWitt equation in quantum cosmology.)
This fact removes the gauge non-invariance
argument since time {\em emerges\/} only for a
particular class of states, and the full quantum
theory is fully invariant. But this approach again
leads to problems: There is no {\em a priori}
reason why the semiclassical states that have a
classical notion of time should be selected as
special. It is usually pointed out that
classicality is necessary in quantum cosmology (a
parameterised system) since we observe a classical
spacetime, but of course this does not constitute
an explanation. Sometimes the notion of
decoherence is evoked, but again this depends on
an {\em a posteriori\/} split between system and
environment that cannot be justified from a first
principles knowledge of quantum theory.
 But the more severe criticism is one that proceeds analogous to
according to that in the classical case: If time
is to be viewed only as a clock
 (semiclassical now rather than classical), the notion of
temporal
 succession is likely to be compromised, for it is not encoded in the
 structure of the clock.

   As we will see this problem is solved when the Dirac
   procedure is implemented in the {\em HPO histories\/} scheme.

   \subsubsection{The quantisation algorithm}
  We shall follow the procedure we have described in section II for
  unconstrained systems. That is, we identify the history group from
a
  study of the space $\Pi$ of classical histories. Then we  seek
   a representation of this group such that the Hamiltonian
   constraint $H_{\kappa}$ is  a well-defined self-adjoint operator.
Then the group of constraints generated
classically by $H_{\kappa}$ for all measurable
functions $\kappa(\cdot)$ is represented in the
history Hilbert space by the unitary operators
  $e^{-i H_{\kappa}}$.

In accordance with our classical analysis, the
group of constraints is a symmetry of the system.
   This means that a projection operator $P$ corresponds to physically
   the same proposition as $e^{iH_{\kappa}}P e^{-i H_{\kappa}}$. This
   correspondence ought to be
   reflected in the assignment of probabilities, and hence in the
   decoherence functional. One has to demand  that
   \begin{equation}
   d(e^{iH_{\kappa}}\alpha e^{-iH_{\kappa}}, e^{i H_{\lambda}} \beta
   e^{-iH_{\lambda}})
   = d(\alpha, \beta)          \label{fund}
   \end{equation}
   for all pairs of projection operators $\alpha, \beta$ and
   measurable functions $\kappa(\cdot), \lambda(\cdot)$.
   \par
   This condition is implemented by substituting $\alpha$ with ${\bf E}
   \alpha {\bf E}$
   in
   the expression for the decoherence functional, where ${\bf E}$ is the
   projector onto the closed linear subspace where
 $H_{\kappa}$ takes values zero for all functions $\kappa(\cdot)$.   Clearly
 the
 property  \ref{fund} is satisfied since $e^{i H_{\kappa}} {\bf E} =
   {\bf E}$ .
This is the natural condition from the perspective
of Dirac quantisation. It implies that only the
gauge-invariant part of the projector is relevant
to the probability assignment. The range of ${\bf
E}$ is the Hilbert space of `physical' histories.
Nonetheless, it is not necessary to restrict our
description on observables living on this space.
Any propositions in ${\mathcal V}$ is acceptable
in the quantum theory. But only its
gauge-invariant part contributes to the
probability assignment.

The projector ${\bf E}$ can be heuristically
considered as the continuous analogue of a tensor
product $\otimes_t E_t$, where $E_t$ is the
projector onto the physical subspace at time $t$.
   In  a discrete time version this amounts to writing the class operator
   that appears
   in the decoherence functional as
   \begin{equation}
   C_{\alpha } = E \alpha_{t_1}(t_1)    E \ldots E \alpha_{t_n}(t_n)E
   = E \alpha_{t_1} E \ldots E \alpha_{t_n}E
   \end{equation}
   Note that the presence of the projectors $E$ implies that the
   Heisenberg-time
   dependence does not affect the final expression.
   \par
   Since ${\bf E}$ remains unaffected when acted upon by $e^{-i
H_{\kappa}}$,
   in the
   final expression of the decoherence functional the result is the
   same as substituting a reduced expression for the ${\cal S}$ operator in
   the decoherence
   functional (together with dropping ${\cal U}$; indeed
   ${\cal U} = e^{-iH_{id}}$  just performs a constraint transformation which
   leaves the
   projector ${\bf E}$ invariant).

   For the operator $\cal S$ we have
   ${\cal S}_{red} = {\bf E} {\cal S} {\bf E}$, or in a discretised form
   \begin{eqnarray}
   {\cal S}_{red} ( | v_{t_1} \rangle \ldots | v_{t_n} \rangle ) =
   E | v_{t_n} \rangle E|v_{t_1} \rangle \ldots E|v_{t_{n-1}} \rangle
   \end{eqnarray}
   Finally, time translations which are represented by the action operator
   $S_{\kappa} = V - H_{\kappa}$, are now to be implemented by
   \begin{equation}
   S_{red} = {\bf E} S_{\kappa} {\bf E} = {\bf E} V {\bf
E}.
   \end{equation}
   \par
   An important property of the Liouville operator is that it leaves
   the decoherence functional invariant. Namely, if we consider the
   transformation
   \begin{equation}
   \alpha \rightarrow \alpha' =e^{isV} \alpha e^{-isV}
   \end{equation}
   then
   \begin{equation}
   d(\alpha', \beta') = d(\alpha, \beta)
   \end{equation}
   A simple proof of this,  follows from the remark that
   the one-parameter group generated by the Liouville operator translates
   the temporal support  of history propositions.
   Hence for a homogeneous
   projector, written formally as $\alpha = \otimes_t \alpha_t$, the
   transformation
   (IV.5) can be thought of
    as
   $\alpha \rightarrow \alpha' = \otimes_t \alpha_{t+s}$. Now the projection
   operator appears in
   the decoherence functional through the class operator $\tilde{C}$. In the
   case of the parameterised system, the class operator does not depend
on the Hamiltonian: rather it
   depends only on the ordering of the single-time projectors $\alpha_t$,
   which is
   not affected by the transformation (IV.5). From this (IV.6) follows.

   This implies that the kinematical time translations are a genuine symmetry
   of the system. They  correspond to an invariance under
   parameterisations of the real axis $\mathR$ of time. Actually, the most
   general symmetries of the quantum system are the quantum version of the
   transformations generated by $V_{\lambda}$.
  The operator $V_{\lambda}$ generates the transformations
  \begin{equation}
   \otimes_{\tau} \alpha_{\tau} \rightarrow \otimes_{\tau} \alpha_{\tau + s}
\end{equation}
where $\tau(t) = \int_a^t ds/\lambda(s)$. Clearly
when $\kappa(t) > 0$, $\tau(t)$ is a strictly
increasing function of $t$, and the ordering is
preserved. Hence with the same argument as before
we get
\begin{eqnarray}
d(e^{isV_{\lambda}}\alpha e^{-isV_{\lambda}},
e^{is V_{\lambda'}} \beta e^{-isV_{\lambda'}}) = d( \alpha, \beta)
\end{eqnarray}
This is the sense in which the probability
assignment is reparameterisation invariant. In the
example given below, this construction will be
made explicit.

   This are the general guidelines for the quantisation procedure of
   parameterised systems. They may need to be modified or augmented in
   individual cases.

    An instance of this is the case where $\xi \neq 0 $. We cannot then take
    the $\xi $
   eigenspace of
   $H_{\kappa}$ as providing the necessary projector to enforce the Dirac
   procedure.
   Rather we should return to a discretised version and consider the
   projectors $E$ at
   the $\xi$-eigenspaces of the single-time constraint operators, and from
   them
   construct a suitable expression for the continuous analogue of ${\bf E} =
   E
   \otimes E \otimes \cdots
   \otimes E$. The projector ${\bf E}$ should again be a spectral projector
   of the Hamiltonian, if
   the invariance condition (IV.1) is to hold. We shall give an example of
   this
   construction later, so we do not further elaborate here.
  \par
We should also remark that a knowledge of the
single-time projectors $E$ is important in the
construction of the decoherence functional.
According to equation (\ref{decfun}) one should consider the
boundary Hilbert space, where the contribution of
the initial and/or final state is contained. This
means that we need consider initial and final
density matrices lying within the physical Hilbert
space as objects entering the definition of 
the operator ${\cal A}$. Thus we have to impose
\begin{equation}
[E,\rho_0] = [E , \rho_f] = 0
\end{equation}
This is equivalent to reducing the boundary
Hilbert space to its physical subspace: the range
of the projector $E \otimes E$.

   \subsection{ The harmonic oscillators with constant energy difference}
   \subsubsection{Canonical treatment}
   The canonical commutation relations
\begin{equation}
[x_i,p_j] = \delta_{ij}
\end{equation}
can be represented in the Fock space
   $e^{\mathC^2}$ on which
Hamiltonian $H = \frac{1}{2}(p_0^2+x_0^2 -p_1^2 -
q_1^2)$ is a well-defined self-adjoint operator.
The creation and annihilation operators are
written in terms of the generators of the
canonical group as
\begin{equation}
b_i = \frac{1}{\sqrt{2}} (x_i + i p_i)
\end{equation}
The Hilbert space contains unnormalised coherent
states $| \exp w \rangle$ where $w = (w_0,w_1) \in
\mathC^2$. The constraint is
\begin{equation}
h = b^{\dagger}_0 b_0 - b^{\dagger}_1 b_1,
\end{equation}
and its eigenstates are clearly
\begin{equation}
|n,m \rangle = \frac{(b^{\dagger}_0)^n}{\sqrt{n!}}
\frac{(b^{\dagger}_1)^m}{\sqrt{m!}} |0 \rangle
\end{equation}
and satisfy
\begin{equation}
h |n,m \rangle = (n - m) |n,m \rangle.
\end{equation}

Clearly, the condition that $h$ takes the value
$\xi$ can be realised only if $\xi $ is an
integer, say $N = n- m $. The $N$-eigenstates of
$h$ are then $|N+m,m \rangle$ for all positive
integers $m$, and the corresponding projector is
\begin{equation}
E = \sum_{m=0}^{\infty} |N+m,m \rangle \langle
N+m,m|.
\end{equation}
Its diagonal matrix elements in a coherent state
basis are
\begin{equation}
\langle \exp w| E|\exp w \rangle =
\sum_{m=0}^{\infty} \frac{1}{\sqrt{n!m!}}
(\bar{w}_0w_0)^{N+m} (\bar{w}_1 w_1)^m
\end{equation}
Let us define
\begin{equation}
e^{I_N[a,b]} = \sum_{m=0}^{\infty}
\frac{1}{\sqrt{(N+m)!m!}} a^{N+m}b^{m}
\end{equation}
Then
\begin{equation}
I_0[a,b] = ab
\end{equation}
and we therefore have
\begin{equation}
\langle \exp w| E|\exp w \rangle =
e^{I_N[\bar{w}_0w_0, \bar{w}_1w_1]}
\end{equation}

\subsubsection{The history space}
For the representation of the history group, we
seek a Hilbert space ${\cal F}$ which can be
written as ${\mathcal V} = \otimes_t H_t$ with
$H_t$ a copy of the Hilbert space of the canonical
theory. The space of smearing functions for the
history algebra can be chosen as the set of
square-integrable, $\mathR^2$-valued functions on
$\mathR$, and its complexification is isomorphic
to $\mathC^2\otimes L^2(\mathR,dt)$. Following the
standard method it is easy to show that
\begin{equation}
{\mathcal V} = e^{ \mathC^2 \otimes
L^2(\mathR,dt)}
\end{equation}
The representation of the history group is
constructed by writing its generators as
combinations of the creation and annihilation
operators
\begin{eqnarray}
x_t^i = \frac{1}{\sqrt{2}} (b^i_t + b^{i
\dagger}_t )\\ p_t^i = \frac{-i}{\sqrt{2}} (b^i_t
- b^{i \dagger}_t)
\end{eqnarray}
It follows that the operator $H_{\kappa}$ exists
as a self-adjoint operator which implements the
automorphisms
\begin{eqnarray}
e^{iH_{\kappa}} b^0_t e^{-iH_{\kappa}} &=&
e^{-i\kappa(t)} b^0_t \\ 
e^{iH_{\kappa}} b^1_t
e^{-iH_{\kappa}} &=& e^{i \kappa(t)} b^1_t
\end{eqnarray}

The Hilbert space ${\mathcal V}$ contains the
natural unnormalised coherent states $|\exp
w(\cdot) \rangle$, with $w(\cdot) \in
\mathC^2\otimes L^2(R,dt)$. This provides the
fundamental relation between the Fock space and
the continuous tensor product of single-time
Hilbert spaces:
\begin{eqnarray}
\otimes_t H_t &=& {\mathcal V} \nonumber \\
\otimes_t | \exp w_t \rangle_{H_t} &\rightarrow &
| \exp w(\cdot) \rangle
\end{eqnarray}

\subsubsection{The spectrum of the Hamiltonian}
According to our previous analysis, in order to
construct the projector onto the physical subspace
we need to study the spectrum of the constraint
operator $H_{\kappa}$. The Hamiltonian has the
generalised eigenstates
\begin{equation}
| t_1, \ldots t_n; t'_1, \ldots, t'_m \rangle :=
\frac{1}{\sqrt{n!m!}} b^{0 \dagger}_{t_1} \ldots
b^{0 \dagger}_{t_n} b^{1 \dagger}_{t'_1} \ldots
b^{1 \dagger}_{t'_m} |0 \rangle
\end{equation}
for which
\begin{eqnarray}
H_{\kappa} | t_1, \ldots t_n; t'_1, \ldots, t'_m
\rangle = \hspace{6cm}   \nonumber \\
 (\kappa(t_1) + \cdots + \kappa(t_n) -
\kappa(t'_1) - \cdots -\kappa(t'_m)) | t_1, \ldots
t_n; t'_1, \ldots, t'_m \rangle
\end{eqnarray}
In terms of the coherent state vectors they read
\begin{eqnarray}
\langle \exp w(.) | t_1, \ldots t_n; t'_1, \ldots,
t'_m \rangle = \frac{1}{\sqrt{n!}} \bar{w}^0(t_1)
\ldots \bar{w}^0 (t_n) \bar{w}^1(t'_1) \ldots
\bar{w}^1(t'_m)
\end{eqnarray}
The actual eigenstates are smeared by real
functions $\phi^{(n,m)} (t_1, \ldots, t_n; t'_1,
\ldots , t'_m) $, that are separately symmetric
with respect to their unprimed and primed
arguments. They are elements of
$(L^2(\mathR,dt)_{\mathR^n})_S \otimes
(L^2(\mathR,dt)_{\mathR^m})_S $. If we denote
\begin{equation}
| \phi^{(r,s)} \rangle = \int dt_1 \ldots dt_n
dt'_1 \ldots dt'_r \phi^{(r,s)} (t_1, \ldots, t_n;
t'_1, \ldots , t'_n) | t_1, \ldots t_s; t'_1,
\ldots, t'_r \rangle
\end{equation}
then we have
\begin{eqnarray}
H_{\kappa} | \phi^{(n,m)} \rangle = \int dt_1
\ldots \int dt_n dt'_1 \ldots dt'_m \nonumber \\
\times (\sum_{i=1}^n \kappa(t_i) - \sum_{i=1}^m
\kappa(t'_i) ) \phi^{(n,m)} (t_1, \ldots, t_n;
t'_1, \ldots, t'_m) | \phi^{(n,m)} \rangle          \label{eigenv}
\end{eqnarray}
The decomposition of unity in terms of the
generalised eigenstates of the constraint operator
reads
\begin{equation}
1 = \sum_{n=1}^{\infty}
\sum_{m=1}^{\infty}\frac{1}{\sqrt{n!m!}} \int dt_1
\ldots dt_n dt'_1 \ldots dt'_m | t_1, \ldots t_n;
t'_1, \ldots, t'_m \rangle \langle t_1, \ldots
t_n; t'_1, \ldots, t'_m |
\end{equation}

\subsubsection{The physical subspace}
Let us now consider the identification of the
projector onto the physical subspace.

\paragraph{The case $\xi =0$.}
In this case we are interested in finding the
subspace of ${\mathcal V}$ on which $H_{\kappa}$
is zero for every $\kappa$. From equation \ref{eigenv} it
is evident that the (generalised) eigenvalues of
the Hamiltonian vanish for all $\kappa(t)$ only
when $n = m$ and $t_i = t_i'$ for all $i = 1,
\ldots n$. This means that the physical subspace
is `spanned' by the generalised eigenvectors of
the form
 $|t_1,\ldots, t_n;t_1, \ldots t_n
\rangle$. The projector ${\bf E}$ is therefore
\begin{equation}
{\bf E} = \sum_{n=1}^{\infty} \frac{1}{n!}\int
dt_1 \ldots dt_n | t_1, \ldots t_n;t_1, \ldots,
t_n \rangle \langle t_1, \ldots t_n;t_1, \ldots
t_n |
\end{equation}
and its diagonal matrix elements in the
coherent-state basis are
\begin{eqnarray}
\langle \exp w(.) | {\bf E} | \exp w(.) \rangle =
\sum_{n=0}^{\infty} \frac{1}{n!} \hspace{5cm} \nonumber \\
\times \int dt_1 \ldots dt_n (\bar{w}^0w^0)(t_1)
\ldots (\bar{w}^0 w^0)(t_n) (\bar{w}^1w^1)(t_1)
\ldots (\bar{w}^1 w^1)(t_n) \nonumber \\ = \exp
\left( \int dt (\bar{w}_0 w_0 \bar{w}_1 w_1)(t)
\right)
\end{eqnarray}
We remind the reader that due to the analyticity
property of the arguments of coherent states, the
knowledge of the diagonal matrix elements of an
operator is sufficient to determine the operator.

\paragraph{The case $\xi \neq 0$.}
As we explained earlier the projection operator
${\bf E}$ does not correspond to the
$\xi$-eigenvalue of the constraint operator.
Rather we should seek to construct
 ${\bf E}$ as  a continuous tensor product of single-time projectors $E_t$
 into the
physical Hilbert space. Each of the single-time
projectors $E_t$ is given by equation (IV.15).
\
From equation (IV.25) we see that the matrix
elements of ${\bf E}$ (a spectral projector of
$H_{\kappa}$) can formally be written as
\begin{equation}
\langle \exp w(.) | {\bf E}| \exp w(.) \rangle =
\prod_t \langle \exp w_t| E_t| \exp w_t \rangle
\end{equation}
and then equation (IV.19) suggests strongly that
the precise form is
\begin{equation}
\langle \exp w(.) | {\bf E}| \exp w(.) \rangle =
\exp \left( \int dt I_N[(\bar{w}_0w_0)(t),
(\bar{w}_1w_1)(t)] \right)
\end{equation}

For $N = 0$ the result is in agreement with
(IV.33) which was obtained using a different line
of reasoning.

\subsubsection{Reparameterisation invariance}
Let us now see how the smeared Liouville operator
appears in the Hilbert space of the system. One
can see that the operator $V_{\lambda} := \sum_i
\int dt \lambda(t) p_{it} \dot{x}_{it}$ generates
the automorphisms
\begin{equation}
e^{isV_{\lambda}} b_w e^{-is V_{\lambda}} = b_{w'}
\end{equation}
where
\begin{equation}
w'(t) := e^{s \kappa(t) \frac{d}{dt}} w(t).
\end{equation}
It is easy to see that the definition
\begin{equation}
\tau_\lambda(t) :=\int_a^t ds/\lambda(s)
\end{equation}
describes the reparameterisation of time, and
gives
\begin{equation}
w'(t) = w(\tau_{\lambda}^{-1}( \tau_{\lambda}(t) +
s))
\end{equation}
in accordance with the classical equation
(III.25).

It is easy to check that the projector ${\bf E}$
remains invariant under the action of
$e^{isV_{\lambda}}$. The decoherence functional is
therefore invariant under reparameterisations of
time.

\section{The parameterised relativistic particle}
The algorithm we gave for the quantisation of
parameterised systems is easily implemented in the
case where the constraint operator has a discrete
spectrum since the physical `subspace' is a
genuine subspace of the Hilbert space. However,
this is not the case if the operator has a
continuous spectrum, and one then has to modify
the algorithm in an appropriate way.

The standard paradigm of a parameterised system
with a continuous spectrum for the constraint
operator is the relativistic particle. In this
section we shall give its classical treatment and
then discuss why the method we employed earlier
for the quantisation needs to be augmented.

 \subsection{ The classical treatment}
 The phase space of this system is $\mathR^4$. It is spanned by the global
coordinates $x_0,x_1,p_0,p_1$, which correspond
respectively to the time
 coordinate, the spatial coordinate, the energy and the momentum of a free
 relativistic particle
in two-dimensional Minkowski spacetime. The phase
space is
 equipped with the
  Poisson bracket
 \begin{equation}
 \{ x_i,p_j \} = \delta_{ij}
 \end{equation}
 where $i,j = 0,1$.
We have the first-class constraint $ h = p_0^2 -
p_1^2 = m^2$, and the Hamiltonian of the system is
proportional to $h - m^2$.

 In the history
 version of this system, a path $\gamma \in \Pi$ is parameterised by the
 coordinates
 $(x_{0t},p_{0t},x_{1t}, p_{1t})$, and the space $\Pi$ carries  a symplectic
 form
 \begin{equation}
 \Omega = \int dt \sum_{i = 0}^1 dp_{it} \wedge
dx_{it}.
 \end{equation}
The corresponding Poisson bracket is
 \begin{equation}
 \{ x_{it}, p_{jt'} \} = \delta_{ij} \delta(t,t'),
 \end{equation}
  and the generator of the symmetry is
 \begin{equation}
 H_{\kappa} = \int dt \kappa(t) h(x_t,p_t).
 \end{equation}
 The action of $H_{\lambda}$ on $\gamma \in \Pi$ is given by
 \begin{equation}
 (x_{0t}, p_{0t}, x_{1t}, p_{1t} ) \rightarrow ( x_{0t}+ p_{0t} \lambda(t) ,
 p_{0t},
 x_{1t} - p_{1t} \lambda(t), p_{1t} )
 \end{equation}

 In the constraint surface (defined by $h(x_t,p_t) = m^2$ for all $t$) one
 can
 use the
 coordinates $(x_{0t}, x_{1t}, a_t)$ where $a_t$ is defined by
 \begin{eqnarray}
 p_{0t} = m \cosh a_t \\
 p_{1t} = m \sinh a_t
 \end{eqnarray}
 and the symplectic form reduces to a degenerate 2-form on $C_h$
 \begin{equation}
 \Omega_{C} = \int m dt \left( \sinh a_t da_t \wedge dx_{0t} + \cosh a_t
 da_t \wedge dx_{1t} \right)
 \end{equation}
 Note that $C_h$ is doubly connected, consisting of a piece with positive and
 one with
 negative energy solutions. For simplicity we restrict our description to the
  former.

The coordinates $\zeta_t = m \sinh a_t x_{0t} + m
 \cosh a_t x_{1t}$
 and $a_t$ are
 easily seen to be constant on the $H_{\kappa}$ orbits, while the coordinates
 $ \xi _t = m \cosh a_t x_{0t} + m \sinh a_t x_{1t}$ correspond to
 degenerate directions of $\Omega_C$. Then $\zeta_t$ and $a_t$ are
 proper coordinates on $\Pi_{red}$,  and the symplectic form $\Omega_{red}$
 is
 obtained
 \begin{equation}
 \Omega_{red} = \int dt\, da_t \wedge d \zeta_t
 \end{equation}
 The action $S_{\kappa}$ acts on elements of $\Pi$ as
 \begin{eqnarray}
 (x_{0t}, p_{0t} , x_{1t}, p_{1t} ) = \hspace{7cm}\nonumber \\
 ( x_{0t+s} + p_{0t+s} \int_0^s \kappa(t +s')
ds',
 p_{0t+s}, x_{1t+s} - \int_0^s \kappa(t+s') ds', p_{1t+s} )
 \end{eqnarray}
 which reduces to the action on $\Pi_{red}$
 \begin{equation}
 (a_t, \zeta_t) \rightarrow (a_{t+s},
\zeta_{t+s}),
 \end{equation}
 and can be represented as
 \begin{equation}
 \tilde{S} = \int dt a_t \dot{\zeta}_t = V_{red}.
 \end{equation}

 The solutions of the classical equations of motion are then all
 elements $\tilde{\gamma}$ of $\Pi_{red}$ such that
 \begin{eqnarray}
 a_t(\tilde{\gamma}) = const. \\
 \zeta_t(\tilde{\gamma}) = const.
 \end{eqnarray}

It is interesting to note the form of the
solutions to the equations of
 motion on $\Pi$. By eliminating the `pure gauge' variable $\xi_t$ we
 can write the relation
 \begin{equation}
 x_{1t} = \frac{ - \cosh^2 a_t x_{0t} + \cosh a_t / m}{1 + \sinh a_t \cosh
 a_t}
 \end{equation}
 For the solutions to equations of motion, $a_t$ and $\zeta_t$ are
 constants. Hence the history variable $- x_{0t}$ can be viewed as a kind of
 clock
 time for the system. This means that, for each $t$, the value of $x_{1t}$
 for a classical
 solution is correlated uniquely to a  value of $x_{0t}$.
 This value can be taken
 as measuring the lapse of time from a chosen origin. This choice of time
 parameter is of course arbitrary and not of physical significance, but
 it conforms to
the standard practice of using the coordinate time
as the time parameter of the system.
 The  correspondence between $x_{1t} $ and $x_{0t}$ along the classical
 solutions is one-to-one at each instant $t$.

\subsection{The need for a non-Fock representation}
According to our previous analysis, the history
Hilbert space for the relativistic particle should
be constructed by considering the representations
of the history group $[\,x^i_t,p^j_{t'}\,] =
i\delta^{ij} \delta(t,t')$ such that the
Liouville and the Hamiltonian operators exist
\begin{eqnarray}
V = \sum_i \int dt\, p^i_t \dot{x}_t^i \\
H_{\kappa} = \int dt\, \kappa(t) (p_{0t}^2 -
p_{1t}^2).
\end{eqnarray}
This implies that the following automorphisms
ought to be implemented unitarily
\begin{eqnarray}
p_{it} &\rightarrow& p_{it} \\ 
x_{0t} &\rightarrow&
x_{0t} + \kappa(t) p_{0t} s \\
 x_{1t} &\rightarrow&
x_{1t} - \kappa(t) p_{1t} s
\end{eqnarray}
Unfortunately there does not exist a Fock
representation on which these automorphisms can be
unitarily implemented.

\subsubsection{Discrete time}
The standard quantum theory for a parameterised
relativistic particle can be readily defined. One
can, for example, use the Schr\"odinger
representation for the canonical group on
$L^2({\bf R^2},dx_0 dx_1)$ in which the constraint
reads
\begin{equation}
h = - \frac{1}{2} \left(\frac{\partial^2}{\partial x_0^2} -
\frac{\partial^2}{\partial x_1^2} \right)
\end{equation}
This has a continuous spectrum. One cannot
therefore write a sharp projector onto the value
$m^2$. It is convenient then to use a regularised
projector $E_{\delta}$ where $\delta $ is a small
number. $E_{\delta}$ is the spectral projector of
$h$ in the range $[m^2 - \delta, m^2 + \delta]$.
In general as $\delta \rightarrow 0$ the matrix
elements of $E_{\delta}$ diverge as $\delta^{-s}$.
\par
Then consider a finite set of $n$ time instants
$\{t_i \}$ $(i = 1, \ldots n) $ and construct the
history Hilbert space ${\mathcal V} =
\otimes_{t_i} H_{t_i}$. Clearly the operator
\begin{eqnarray}
e^{-iH_{\kappa}} = \otimes_{t_i} e^{-i h_{t_i}
\kappa(t_i)}
\end{eqnarray}
exists (it is the discrete version of the operator
${\cal U}$ introduced in \ref{Heisen_op} . We can also
construct the projection operator
\begin{equation}
{\bf E}_{\delta} = \otimes_{t_i} E_{\delta t_i}
\end{equation}
As explained earlier in Section IV, the projection
onto the physical subspace is equivalent to
substituting a reduced Schr\"odinger operator in
the decoherence functional. Hence we can write a
regularised reduced Schr\"odinger operator as
   \begin{eqnarray}
  {\cal S}_{\delta}( | v_{t_1} \rangle \ldots | v_{t_n} \rangle ) =
   E_{\delta} | v_{t_n} \rangle E_{\delta}|v_{t_1} \rangle \ldots
   E_{\delta}|v_{t_{n-1}} \rangle
   \end{eqnarray}
One can therefore write a regularised version of
the decoherence functional $d^{\delta}(\alpha,
\beta)$
 substituting ${\cal S}_{\delta}$ in equation \ref{decfun}.
   As we said earlier, the matrix elements of $E_{\delta}$ diverge as
   $\delta^{-r}$  when
   $\delta
   \rightarrow 0$, and hence the renormalised ${\cal S}$ should be defined as
   the
   weak limit
   of
   \begin{eqnarray}
   {\cal S}_{ren}       ( | v_{t_1} \rangle \ldots | v_{t_n} \rangle ) =
    \lim_{\delta \rightarrow 0} \delta^{rn} \left( E_{\delta} | v_{t_n}
    \rangle E_{\delta}|v_{t_1} \rangle
   \ldots E_{\delta}|v_{t_{n-1}} \rangle \right)
   \end{eqnarray}
This is equivalent to redefining the decoherence
functional for discrete times as
\begin{equation}
d_{ren}(\alpha, \beta) =C \lim_{n \rightarrow
0} \delta^{2rn} d^{\delta} (\alpha,
\beta )
\end{equation}
This is a finite, generally non-zero number.
 The coefficient $C$ is introduced here to secure the normalisation condition
 $d(1,1) = 1$.
\par
We can therefore define a decoherence functional
for all discrete time histories.

    The situation for continuous times is very
different since it is not in general possible to
define a {\em continuous\/} tensor product of a
family of operators. Instead, one needs to employ
a more sophisticated approach to the problem, and
this will be the content of a later paper
\cite{AIS00}.

\section{Conclusions}
We have shown that the continuous histories
programme can accommodate the description of
constrained systems using a variant of the Dirac
method. In particular, we have seen how the use of
an HPO history theory avoids the problem of time
that appears classically and quantum mechanically
in the canonical treatment. This is a major reason
for using the history methods to study systems of
this type.

However, there are two distinct issues that remain
to be addressed. The first is of a technical
nature: extend the quantisation algorithm to deal
with systems that have constraints with continuous
spectra. This involves finding physically
meaningful representations of the appropriate
history algebra: these are expected to be not of
Fock type \cite{AIS00}.

The other issue is of a conceptual type and is
related to the interplay between general
covariance and the causal structure in general
relativity. We have said that histories are viewed
as objects that have an intrinsic temporal
ordering. On the other hand, in general relativity
the causal structure is obtained {\em after\/}
solving the equations of motion. One might
question therefore whether a history description
of the theory necessarily violates general
covariance.

Indeed this touches in a second, more general,
aspect of the problem of time in quantum gravity.
How does one know that the spacelike foliation
with respect to which the quantum theory is
defined remains spacelike at the next time step if
the evolution law is not deterministic? We believe
we will have a meaningful answer to this question
when
 we examine the implementation of spacetime
diffeomorphisms in the histories
 version of general relativity.
This is something we intend to address soon. Here,
let us note only that our formalism involves a
mixture of both Lagrangian and Hamiltonian
techniques. As such it allows both the
 group of spacetime diffeomorphisms,
and the group of constraints, to be represented on
the history space by symplectic transformations.
For this reason we expect to be able to clarify
the relation of these symmetries.

We conclude with two remarks. First, in the simple
systems studied in this paper, general covariance
is identical to reparameterisation invariance.
Indeed, the smeared Liouville operator is the
generator of the most general order-preserving
diffeomorphisms of the real line. Diffeomorphisms
are then structurally distinct from the
Hamiltonian constraints and we would expect them
to be implemented as transformations on the
reduced phase space of general relativity.

Second, the problem of time in the quantisation of
general relativity is conceptually distinct from
the problem of time in the quantisation of
parameterised systems. The latter is an artifact
of the canonical approach and, as we have seen,
can be solved by using the histories quantisation
method. Only a rather conservative modification of
the standard quantum mechanical formalism was
required to do this. However, the problem of time
in quantum gravity is grounded in a deep disparity
between the notion of time in quantum theory and
general relativity. Quantum theory seems to
require an {\em a priori} notion of causality,
while in general relativity causality emerges from
the implementation of the dynamics. The resolution
of this issue is likely to require a much more
radical change of the quantum mechanical formalism
than has been proposed so far.

\section*{Acknowledgements}
We would like to thank Chris Isham for useful discussions and for his active assistance on the writing of this paper.
K.S. was supported by a gift from the Jesse Phillips Foundation and C.A. from the NSF grant PHY98-00967.


\begin{thebibliography}{99}

\bibitem {Gri84} R. B. Griffiths.
  \newblock Consistent Histories and the
 Interpretation of Quantum Mechanics.
\newblock {\em  J. Stat. Phys.}  36: 219 (1984).

\bibitem {Omn8894} R. Omn\`es.
\newblock  Logical Reformulation of Quantum Mechanics: I
Foundations.
\newblock {\em J. Stat. Phys.}   53: 893 (1988);
\newblock  The Interpretation of Quantum Mechanics.
\newblock {\em Princeton University Press, Princeton}, 1994 .

\bibitem {GeHa9093} M.\ Gell-Mann and J.\ B.\ Hartle.
 \newblock  Quantum mechanics in
the Light of Quantum Cosmology.
\newblock  In {\em
Complexity, Entropy and the Physics of
Information}, edited by W.\ Zurek.
\newblock  {\em  Addison
Wesley, Reading}, 1990; 
\newblock  Classical Equations
for Quantum Systems.
\newblock {\em Phys. Rev.}   D47: 3345, 1993.

\bibitem{Har93a}
J. B. Hartle.
\newblock Spacetime quantum mechanics and the quantum mechanics of spacetime.
\newblock In {\em Proceedings on the 1992 Les Houches School, 
	Gravitation and Quantisation}. 1993.

\bibitem{I94}
C.J. Isham.
\newblock Quantum  logic and the histories approach to quantum theory.
\newblock {\em J. Math. Phys. } 35:2157, 1994.

\bibitem{IL94}
C.J. Isham and N.~Linden.
\newblock Quantum temporal logic and decoherence functionals in the histories
approach to generalised quantum theory.
\newblock {\em J. Math. Phys. } 35:5452, 1994.

\bibitem{IL95}
C.J.Isham and N.~Linden.
\newblock
Continuous histories and the history group in
generalised quantum theory.
\newblock {\em J. Math. Phys. } 36: 5392, 1995

 \bibitem{ILSS98} C. Isham, N. Linden, K. Savvidou and S. Schreckenberg.
\newblock  Continuous time and consistent histories.
\newblock  {\em J. Math. Phys.} 37:2261, 1998.

\bibitem{Sav99} K. Savvidou.
 \newblock The action operator in continuous time
histories.
\newblock  {\em J. Math. Phys.}  40: 5657, 1999.

\bibitem{An00} C. Anastopoulos. 
\newblock  Quantum Fields in Nonstatic background: a
Histories Perspective.
\newblock  {\em J. Math. Phys.}  February 2000.
gr-qc/ 9903026.


\bibitem{I92} C. J. Isham.
\newblock   Canonical Quantum gravity and the Problem
of Time.
\newblock In {GIFT Seminar 1992}:0157-288.
gr-qc/9210011.

\bibitem{Kuc91} K. Kuchar. 
\newblock  Time and Interpretations of Quantum Gravity
\newblock {\em Winnipeg 1991, Proceedings, General Relativity and
Relativistic Astrophysics}, 211.

\bibitem{Sav99b} K. Savvidou. 
\newblock 
 Continuous Time in Consistent Histories.
\newblock gr-qc/9912076.

\bibitem{A60}
H.~Araki.
\newblock Hamiltonian formalism and the canonical commutation
relations in quantum field theory.
\newblock {\em J. Math. Phys. } 1:492, 1960.


\bibitem{I83} C. J. Isham.  
\newblock Topological and Global
Aspects of Quantum Theory.
\newblock  In {\em Proceedings of the 1983 Les Houches School, Relativity,
Groups and Topology II},  1983.

\bibitem{Rov90} C. Rovelli.
 Quantum Mechanics without Time: a Model.
\newblock {\em Phys. Rev.}  D42: 2638, 1990.

\bibitem{Rov91} C. Rovelli. 
\newblock  Time in Quantum Gravity: a Hypothesis.
\newblock {\em  Phys.
Rev.}  D43: 442, 1991.

\bibitem{Har96} J. B. Hartle. 
\newblock  Time and Time Functions in Parametrized
Nonrelativistic Quantum Mechanics. 
\newblock {\it Class. Quant.
Grav.}  13: 361, 1996.

\bibitem{AIS00} C. Anastopoulos, C. Isham and K. Savvidou. 
\newblock  Histories quantisation of parameterised systems: II. The relativistic particle.
\newblock  Work in preparation.
 
\end{thebibliography}
\end{document}